\documentclass{webofc}
\usepackage{xcolor}
\usepackage[varg]{txfonts}   
\usepackage{hyperref}
\hypersetup{
    colorlinks=true,
    linkcolor={red!50!black},
    citecolor={red!50!black},
    urlcolor={blue!80!black}
}
 
\begin{document}

\title{Quark matter and nuclear astrophysics: recent developments}

\author{\firstname{Tyler} \lastname{Gorda}\inst{1}\fnsep\thanks{\email{gorda@itp.uni-frankfurt.de}} }

\institute{Institut f\"ur Theoretische Physik, Goethe Universit\"at,   Max-von-Laue-Str. 1, 60438 Frankfurt am Main, Germany}

\abstract{%
Does deconfined cold quark matter occur in nature? This is currently one of the fundamental open questions in nuclear astrophysics.  
In these proceedings, I review the current state-of-the-art techniques to address this question in a model-agnostic manner, by synthesizing inputs from astrophysical observations of neutron stars and their binary mergers, and first-principles calculations within nuclear and particle theory.
I highlight recent improvements in perturbative calculations in asymptotically dense cold quark matter, as well as compelling evidence for a conformalizing transition within the cores of massive neutron stars.
}
\maketitle
 
\noindent The question of whether deconfined cold (zero temperature) quark matter (QM) exists in the cores of massive neutron stars (NSs) is one of the fundamental open questions facing nuclear astrophysics today \cite{GWinitiative:2020}. 
Due to the breakdown of lattice Monte-Carlo techniques at large net baryon densities $n_\mathrm{B}$ \cite{deForcrand:2009zkb,Nagata:2021ugx}, direct lattice simulations of Quantum Chromodynamics (QCD) are not available to study these objects.
Rather, addressing this question requires a framework for synthesizing astrophysical data about the bulk properties of NSs \cite{Antoniadis:2013pzd,Cromartie:2019kug,Fonseca:2021wxt,Miller:2019cac,Riley:2019yda,Miller:2021qha,Riley:2021pdl, TheLIGOScientific:2017qsa} with all the currently available data on dense strongly interacting matter, from nuclear experiments at low densities through to particle-theory experiments at very high center-of-mass energies. 
Furthermore, it requires the development of criteria for the identification of microphysical properties of dense matter from its thermodynamic properties \cite{Annala:2019puf,Fujimoto:2022ohj,Annala:2023cwx}.

Nuclear and particle data can be used as input to state-of-the-art chiral effective field theory \cite{Tews:2012fj,Lynn:2015jua,Drischler:2017wtt,Drischler:2020hwi,Keller:2022crb} and perturbative QCD (pQCD) \cite{Kurkela:2009gj,Kurkela:2016was,Gorda:2018gpy,Gorda:2021znl,Gorda:2021kme,Gorda:2021gha,Gorda:2023mkk} calculations in their regimes of validity. 
These calculations can then be used within a NS equation-of-state (EOS) inference setup, combined with the astrophysical observations as well as a chosen prior model for describing the NS EOS. 
Many groups now use such a model-agnostic framework for constraining the NS EOS, as it allows one to deduce the properties of dense matter without introducing a large number of modeling assumptions.
The main assumption is form of the prior model for the NS EOS used in the analysis.

In the last two years, there has been a particular focus in the literature on the role of the pQCD constraint within these setups \cite{Gorda:2022jvk,Somasundaram:2022ztm,Essick:2023fso,Brandes:2023hma,Mroczek:2023zxo}.
This discussion has been driven by the work \cite{Komoltsev:2021jzg}, which introduced a clear understanding of how thermodynamic information at $n_\mathrm{B} \gtrsim 40 n_\mathrm{sat}$, with $n_\mathrm{sat} \approx 0.16$~fm$^{-3}$ the nuclear saturation density, impacts the EOS below the $5-7n_\mathrm{sat}$ that are probed in the cores of massive NSs. 
This work showed that the requirements that the NS EOS be causal, stable, and thermodynamically consistent allows specific extreme EOSs at $2.2-2.5n_\mathrm{sat}$ to be ruled out by the location of the pQCD EOS at high densities. 
That is, the pQCD EOS alone begins to rule out EOSs with very high pressures at low densities, even before the inclusion of astrophysical constraints.

These proceedings detail recent improvements to the pQCD EOS used within the analysis of the NS EOS, as well as recent developments in addressing the question of whether the cores of massive NSs likely contain QM or not.

\begin{figure}[t]
\centering
    \raisebox{0.6cm}{\includegraphics[width=0.20\textwidth,clip]{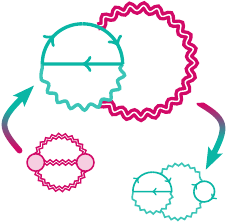}}
$\;$
\includegraphics[width=0.38\textwidth,clip]{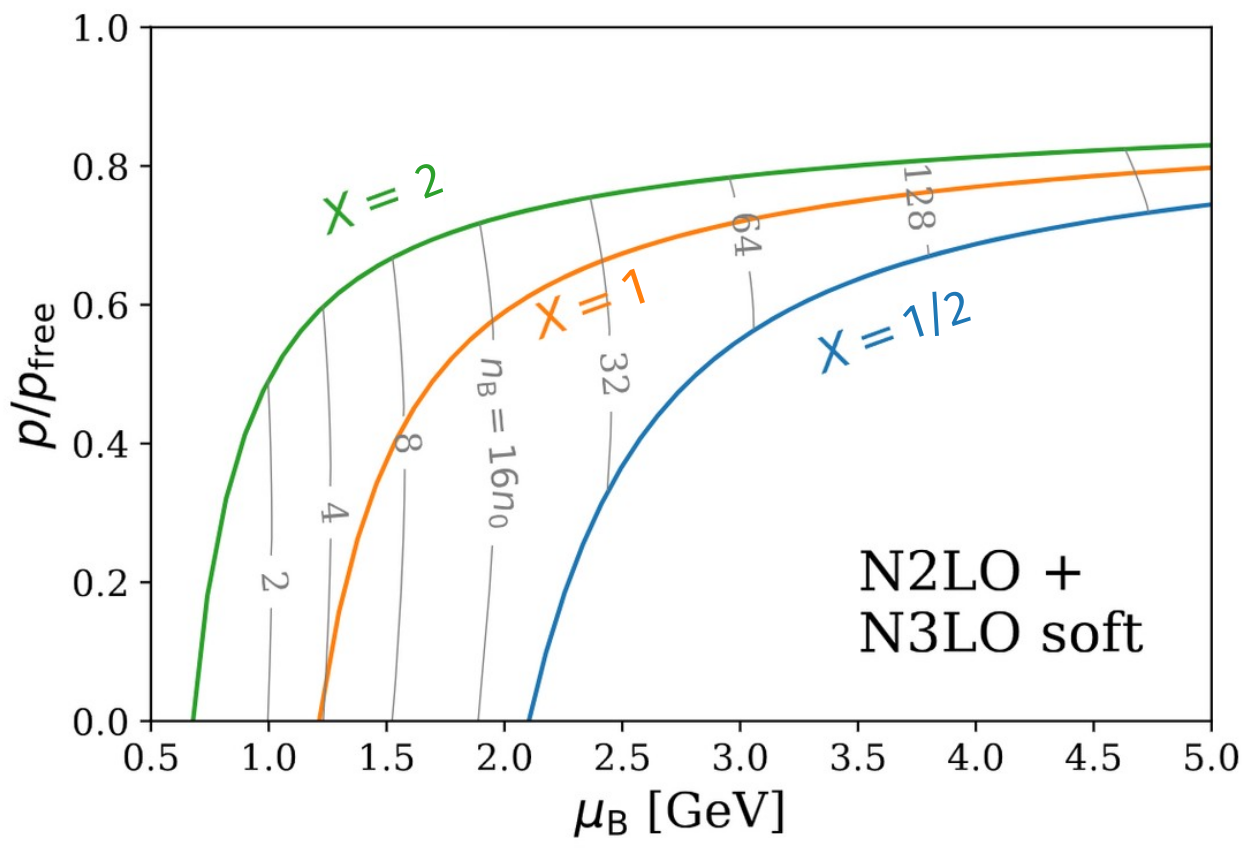}
$\;$
\includegraphics[width=0.38\textwidth,clip]{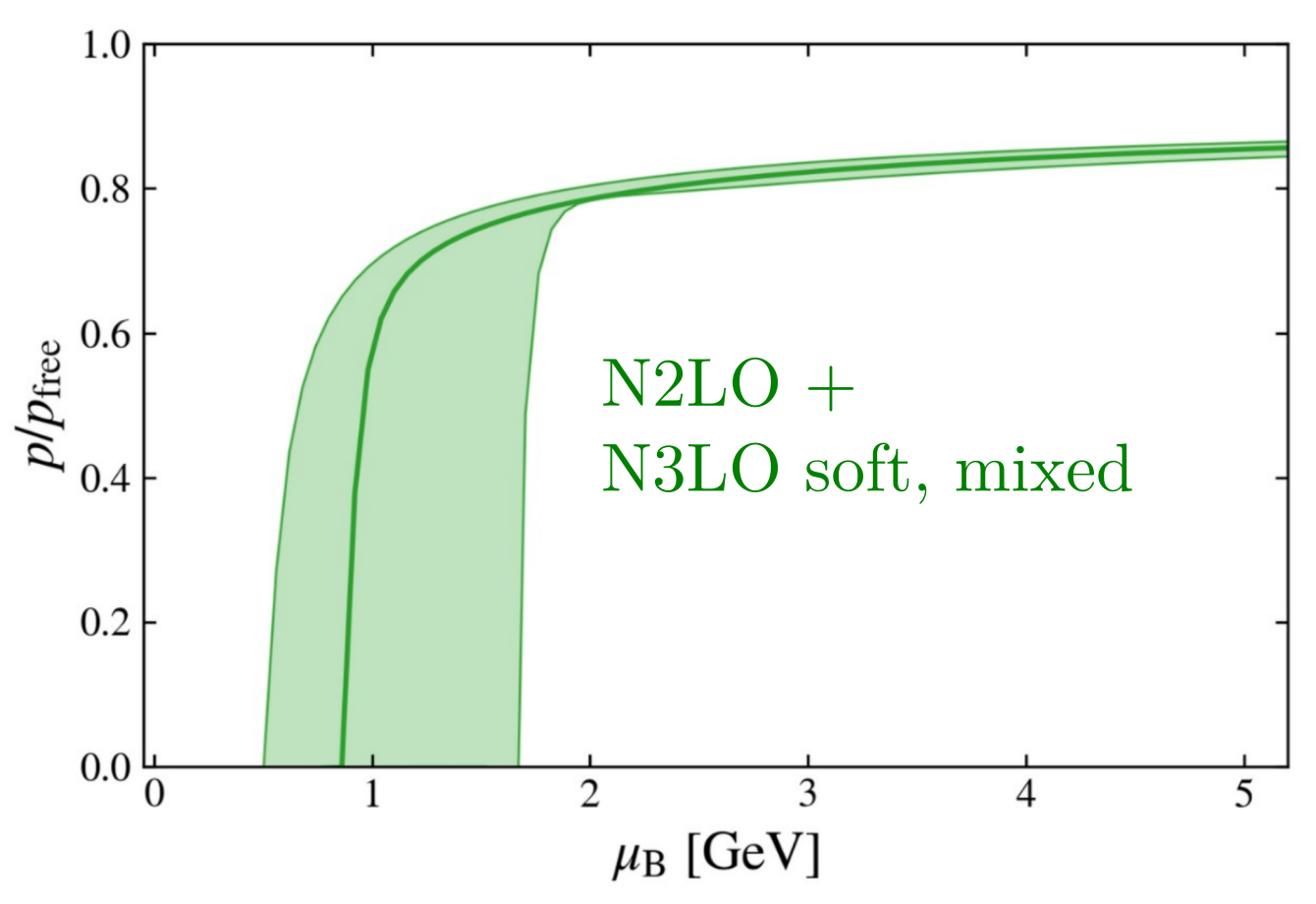}
    \caption{Left: Sample diagrams from the different kinematic contributions to the pressure at $O(\alpha_s^3)$ in pQCD. 
    Clockwise from bottom left, the three figures denote the soft, mixed, and hard contributions. 
    Here, the red lines and vertices denote hard-thermal/dense-loop propagators and vertices respectively. 
    Center: The pressure normalized by the free pressure as a function of baryon chemical potential $\mu_\mathrm{B}$ from the results in \cite{Gorda:2021znl}. 
    Also shown in the figure are contours of constant baryon density. 
    Right: The pressure normalized by the free pressure as a function of $\mu_\mathrm{B}$ including all but the hard sector at $O(\alpha_s^3)$.}
\label{fig-1}     
\end{figure}

\section{Improvements to the pQCD EOS}
\label{sec:pQCD}

In the last five years, the analysis of the pQCD EOS has been pushed towards $O(\alpha_s^3)$ in the strong coupling constant $\alpha_s$. 
In particular, in 2021, the overall structure of the EOS at this order was made clear \cite{Gorda:2021kme} via an organization of Feynman diagrams into classes based on the number of long-wavelength, dynamically screened gluonic loops (see Fig.~\ref{fig-1}, left). 
The fully \emph{soft} sector at this order---consisting of hard-thermal/dense-loop diagrams with two screened gluon loops---was computed in \cite{Gorda:2021znl,Gorda:2021kme}, and the result for the pressure is shown in the center panel of Fig.~\ref{fig-1}. 
In this panel, one can see the residual renormalization-scale dependence of the result, in terms of the parameter $X \equiv 3\overline{\Lambda} /(2 \mu_\mathrm{B})$, with $\overline{\Lambda}$ the renormalization scale in the $\overline{\mathrm{MS}}$ scheme and $\mu_\mathrm{B}$ the baryon chemical potential. 
By estimating the truncation error of this pQCD result by the residual renormalization-scale dependence at this order (for this variation is $O(\alpha_s^4)$), \cite{Gorda:2021znl} found that the result is well converged down to around $n_\mathrm{sat} \approx 40n_\mathrm{sat}$.

Very recently, the calculation of the \emph{mixed} sector at $O(\alpha_s^3)$---consisting of corrections within the hard-thermal/dense-loop theory---has been computed in \cite{Gorda:2023mkk}, bringing the zero-temperature pQCD calculation on par with its high-temperature counterpart from 2003 \cite{Kajantie:2002wa}. 
The result shows a remarkably small dependence on the renormalization-scale parameter $X$ (see Fig.~\ref{fig-1} right), indicating that the screened gluonic sector appears to be under good perturbative control up to this order in perturbation theory. 
The only remaining contributions at $O(\alpha_s^3)$ originate from the \emph{hard} sector, consisting of (potentially infrared-divergent) four-loop diagrams in full QCD. 
These diagrams will be technically challenging to compute, and may face subtleties associated with strictly zero-temperature loop calculations \cite{Gorda:2022yex}, but there are indications from \cite{Gorda:2023mkk} that the full $O(\alpha_s^3)$ result may remain well converged down to $n_\mathrm{B} \approx 27n_\mathrm{sat}$.

\begin{figure}[t]
\centering
    \raisebox{0.05cm}{\includegraphics[height=3.85cm,clip]{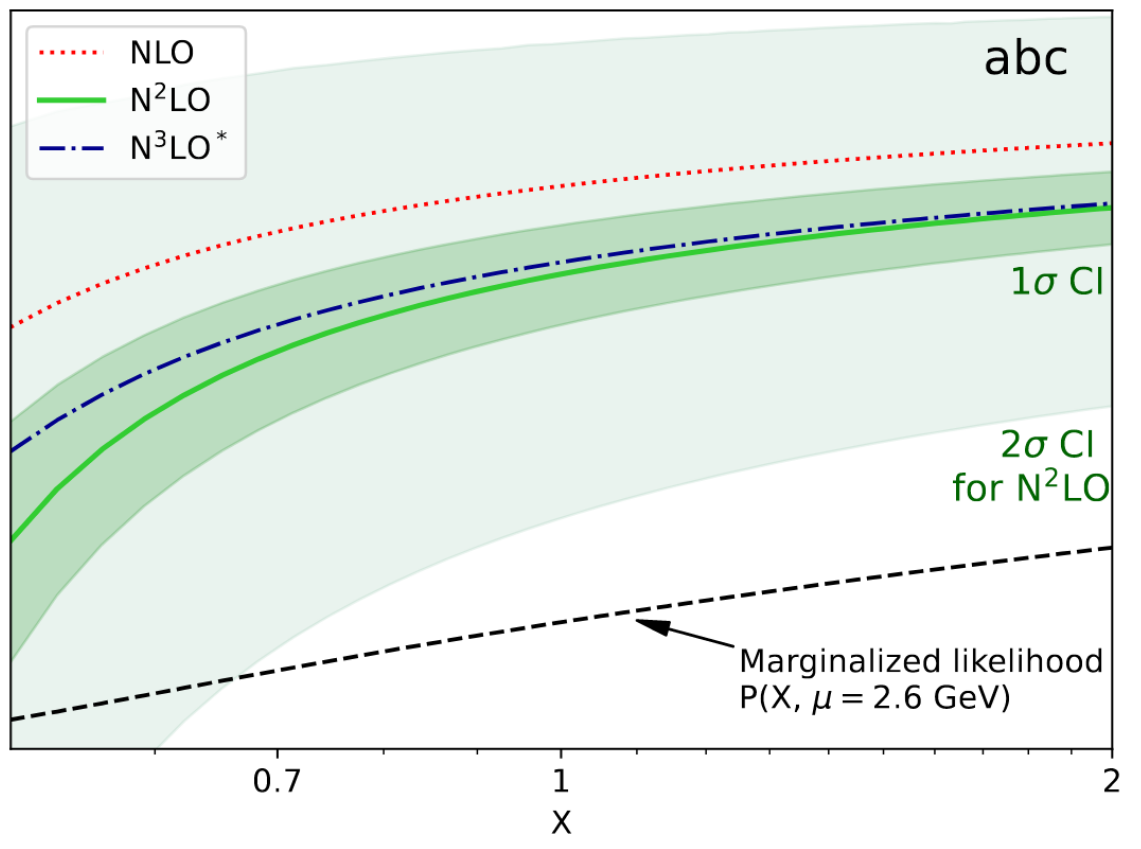}}
$\qquad$
\includegraphics[height=3.95cm,clip]{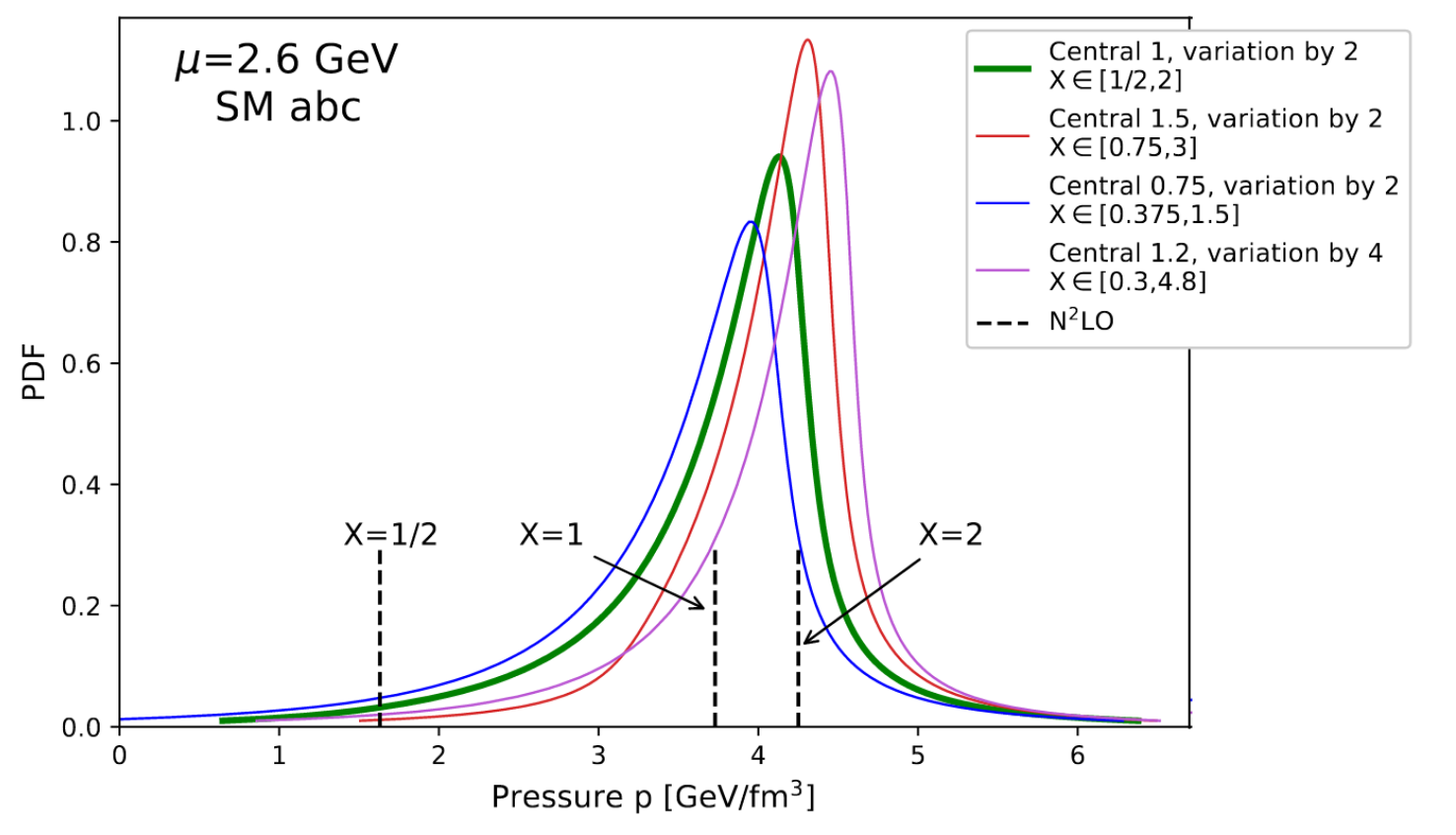}
    \caption{Left: A Bayesian estimate of the $O(\alpha_s^3)$ term of the pQCD series (N$^3$LO) at high density after training on the $O(\alpha_s)$ (NLO) and $O(\alpha_s^2)$ (N$^2$LO) terms using the $abc$ model of convergent series \cite{Duhr:2021mfd}. 
    The result is shown as a function of $X$, and the marginalized likelihood (evidence) for 
    the different $X$ values are shown as a dashed black line (with arbitrary overall normalization). 
    This indicates that larger values of $X$ show better convergence within the model. 
    Shown in the dashed-dotted blue line is the partial $O(\alpha_s^3)$ results of \cite{Gorda:2021znl}.
    Right: A probability density function of the $O(\alpha_s^3)$ pressure of cold QM at $\mu_\mathrm{B} = 2.6$~GeV using the scale-marginalization (SM) prescription and the $abc$ model. 
    The SM procedure conducts an integral of the left figure over the different $X$ values with a logarithmic scale in $X$ and weights each $X$ by its marginalized likelihood.}
\label{fig-2}       
\end{figure}

Another recent improvement in the pQCD EOS comes from a Bayesian uncertainty estimation of the truncation errors \cite{Gorda:2023usm}. 
As mentioned above, the historical way to estimate the error arising from truncating the perturbative series has been to vary the renormalization scale by a factor of two about some central value. 
However, while this does capture the explicit $X$ dependence of the next term in the pQCD series, it does not in any way estimate the $X$-independent part of the remaining terms.
Moreover, it is not clear how to attach a statistical meaning to the error band returned by this procedure.

The work \cite{Gorda:2023usm} instead used a machine-learning based Bayesian interpretation of the uncertainties, in which the perturbative series for the pressure at fixed $\mu_\mathrm{B}$ and $X$ is modeled as draws from a statistical model of convergent series \cite{Duhr:2021mfd}, trained with the available terms using Bayes's theorem
\begin{equation}
    P(\vec{p} \,|\, \text{data}) = \frac{P(\text{data} \,|\, \vec{p}) P_0(\vec{p})}{\int \mathrm{d}{\vec{p}}\, P(\text{data} \,|\, \vec{p})P_0(\vec{p}) }
\end{equation}
Here, $\vec{p}$ represents the parameters of the model, $P_0(\vec{p})$ is a prior distribution of these parameters \cite{Duhr:2021mfd}, $P(\text{data} \,|\, \vec{p})$ is the likelihood that the model with parameters $\vec{p}$ reproduces the data, and the denominator is the marginalized likelihood or evidence of the data given the model and is usually written $P(\text{data})$. 
In the case of \cite{Gorda:2023usm}, the data are the terms in the series for the pressure $p$ normalized by the free pressure $p_\mathrm{free}$ up to $O(\alpha_s^2)$.
The trained model then returns a probability density function for the $O(\alpha_s^3)$ term in the series for each fixed $\mu_\textrm{B}$ and $X$ (see Fig.~\ref{fig-2}, left), which is an estimate for the sum of the series.

Since the dependence on the parameter $X$ is unphysical, and would not remain if one had an infinite number of terms in the series, it is desirable to incorporate the $X$ dependence into the result and return a distribution for the pressure itself.
There are two ways to do this, depending on whether one considers $X$ as an index labeling independent predictions or instead as a hidden parameter.
In the former case, one could average over the predictions for different $X$ values, with a log-uniform prior (for $X$ in some fixed range) $P_0(X)$, while in the latter case, one would additionally fold in the evidence for different $X$ values and marginalize
{\allowdisplaybreaks
\begin{align}
    P(p / p_\text{free} \,|\, \text{data}) &= \int dX\, P\bigl(p(X) / p_\text{free} \,|\, \text{data}(X)\bigr) P_0(X), \;\;\qquad\quad (\text{scale-averaging})\\
    P(p / p_\text{free} \,|\, \text{data}) &= \int dX\, P\bigl(p(X) / p_\text{free} \,|\, \text{data}(X) \bigr) P\bigl(X \,|\, \text{data}(X)\bigr). \;\; (\text{scale-marginalization})
\end{align}
}%
The right panel of Fig.~\ref{fig-2} shows the distribution in the scale-marginalization prescription for different choices of the $X$ interval. Generically, one finds that larger pressures are favored.

These distributions can then be used in a Bayesian inference of the NS EOS.
This has also been performed in \cite{Gorda:2023usm}, which concluded that this analysis leads to only minor differences in the resulting posterior NS EOS distributions, increasing confidence in the use of the pQCD input within these NS EOS inference setups.

\begin{figure}[t]
\centering
\includegraphics[width=0.45\textwidth,clip]{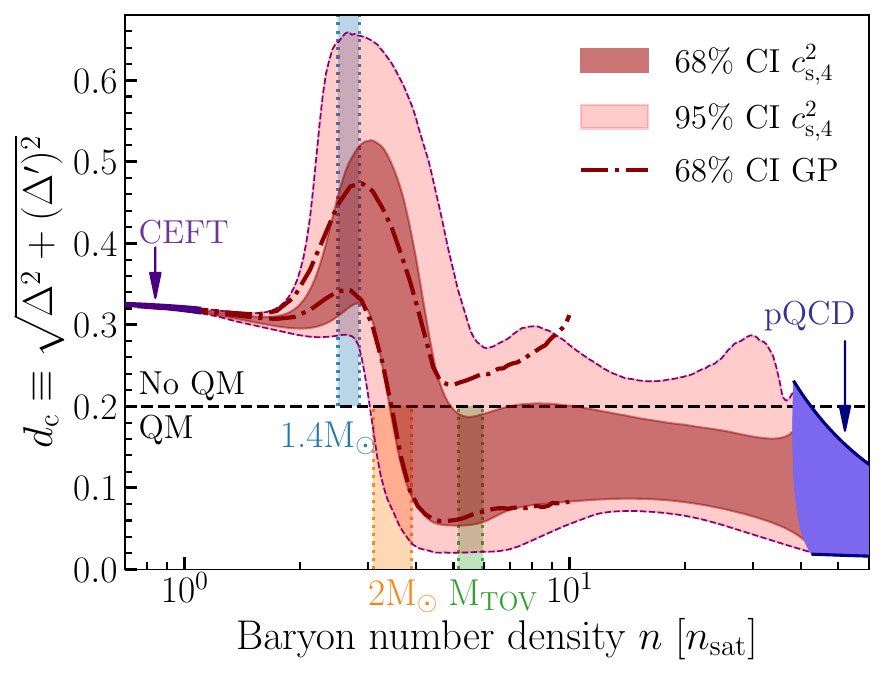}
$\qquad$
\includegraphics[width=0.48\textwidth,clip]{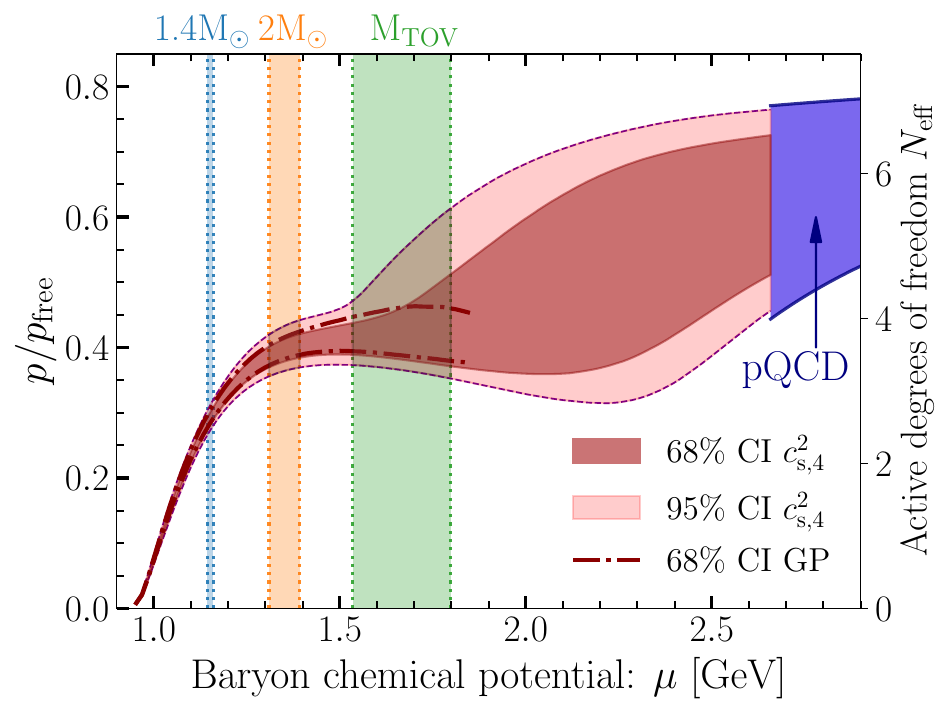}
    \caption{Left: The degree of conformality of NS matter as a function of baryon density. 
    Plotted are 68\% and 95\% credible intervals of the quantity $d_c$. 
    Shown in colored shaded regions are 68\% credible intervals for the central density of stars of different masses. 
    Right: The pressure normalized by the free pressure as a function of baryon chemical potential.}
\label{fig-3}      
\end{figure}

\section{Non-conformal to conformal transition at high densities}
\label{sec:dc_criterion}

A key point in addressing whether QM is located in the cores of massive NSs is the definition of a criterion for delineating the microphysical phase of NS matter based on its thermodynamic properties. 
Such a delineation was first attempted in \cite{Annala:2019puf}, where an analysis of the speed of sound $c_s^2$, polytropic index $\gamma \equiv \mathrm{d} \log p / \mathrm{d} \log \epsilon$ with $\epsilon$ the energy density, and $p/p_\mathrm{free}$ was performed.
This work found evidence for a conformalizing transition along the stable NS sequence and defined the approximate criterion $\gamma < 1.75$ for the presence of QM.
Last year, \cite{Fujimoto:2022ohj} suggested an analysis of the trace anomaly $\Delta \equiv 1/3 - p/\epsilon$ as an even more promising direction.
For a conformal system, $\Delta = 0$.
However, non-conformal systems can still briefly obtain $\Delta = 0$ at a particular density if $\Delta' \equiv \mathrm{d} \log \Delta / \mathrm{d} \log \epsilon \neq 0$.
In this case, the non-conformal system would briefly pass through the value $\Delta = 0$, but not remain there to higher densities, which would be necessary for true conformality.
In fact, at high densities, many microphysical models of dense NS matter exhibit small values of $\gamma$ and $\Delta$ even while remaining in the non-conformal hadronic phase \cite{Annala:2023cwx}. 

In \cite{Annala:2023cwx}, the authors introduce a non-local measure of conformality, namely $d_c \equiv \sqrt{\Delta^2 + (\Delta')^2}$.
This quantity is also zero in a conformal system, but will be large in a generic non-conformal system, even when $\Delta = 0$.
The authors chose a dividing line between non-conformal and conformal matter to be $d_c = 0.2$, as this constitutes the approximate average between the value of the quantity within chiral effective field theory and pQCD in their respective regimes of convergence.
As shown in the left panel of Fig.~\ref{fig-3}, $d_c$ exhibits a remarkably clear change of behavior around $3-4n_\mathrm{sat}$, below the maximum densities reached in massive NSs, signaling a conformalizing  transition along the stable NS sequence.
As analyzed in detail in \cite{Annala:2023cwx}, this change in behavior of NS matter results from a combination of the astrophysical inputs and the pQCD constraint at high densities.
Without the pQCD constraint, such a conformalizing transition is not clearly seen.
Moreover, this transition was found to be robust to the choice of prior model for the NS EOS \cite{Annala:2023cwx}.

This work also analyzed the effective number of degrees of freedom of the matter, in the form of $p/p_\mathrm{free}$---a quantity which is used at high temperature in the analysis of lattice data.
As shown in the right panel of Fig.~\ref{fig-3}, $p/p_\mathrm{free}$ exhibits a flattening within the cores of the most massive NSs, indicative of an approximately conformal system.
Moreover, the quantity takes a value that is about 2/3 of the value in weakly interacting cold QM at high densities, suggesting that this approximately conformal phase may indeed be identified with QM, as would be natural within the context of dense strongly interacting matter.

\section{Outlook}

As detailed above, there is compelling evidence for a conformalizing transition along the stable NS sequence once the current state-of-the-art pQCD input is used, suggesting that cold QM lies in the cores of the most massive stable NSs. 
One important caveat to this is if the NS EOS exhibits a strong phase transition. 
In this case, the EOS can remain non-conformal up to the maximum density reached in NSs \cite{Annala:2019puf,Annala:2023cwx}.
In this context, it would be important to conduct future analyses with explicit phase transitions in the NS EOS prior, as these permit more general EOS behavior that is still consistent with astrophysical observations \cite{Gorda:2022lsk}.

Additionally, it is desirable to find other lines of evidence beyond this thermodynamic analysis that supports or contradicts the hypothesis of QM in the cores of massive NSs.
An analysis of NS cooling or transport properties within the core of NSs would give a much more direct window into the character of low-energy excitations in these objects.
It would also be important to systematically investigate whether the presence of QM has an impact on the post-merger phase of a binary NS merger, or whether there are other clear observables of its presence, such as suggested by \cite{Casalderrey-Solana:2022rrn}.

\bibliography{references.bib}

\end{document}